\documentclass[amsmath,twocolumn,showpacs,aps,prl]{revtex4}
\usepackage{bm}
\usepackage{graphics}
\usepackage{graphicx}
\begin{document}
\title{Dynamic conductivity in graphene beyond linear response}
\author{E.G.~Mishchenko}
\affiliation{Department of Physics and Astronomy, University of Utah, Salt Lake
City, Utah 84112, USA}

\begin{abstract}
The independence of the dynamic conductivity of
intrinsic graphene of frequency takes its origin in the compensation
of the vanishing density of states by the diverging matrix element
of the corresponding interband transition. The applicability of the linear
response approach, however, breaks down when this matrix element
becomes comparable with the inverse electron lifetime. We show that
the physics of the ac conductivity in this regime is determined by Rabi oscillations and obtain it beyond the first order perturbation
theory. Under strong applied electric fields the induced current
eventually saturates at a value determined by the frequency and the
lifetime. We also calculate the electromagnetic response of a graphene
sheet and find that the optical transparency is increased by the
non-linear effects and make experimental predictions.
\end{abstract}


\maketitle

{\it Introduction.} Low-temperature dynamic conductivity ($k_BT \ll
\hbar \omega$) of graphene was predicted to be of the order of the
conductance quantum and independent of frequency,  $\sigma_0
=e^2/4\hbar$ \cite{ludwig}, as long as the frequency $\omega$ is
small compared with  the electron  bandwidth \cite{ryu}. This prediction has
been verified experimentally \cite{li, kuzmenko, mak}. Theoretical
attempts to understand the role played by electron-electron
interaction began even earlier \cite{2}. Following the conclusion
that the interactions do not modify the conductivity in the limit
$\omega \to 0$ \cite{sheehy,herbut} it was further predicted that
even at non-vanishing frequencies the interaction corrections are
extremely small \cite{mishchenko,sheehy2} and should hardly be
noticeable at all, the claim consistent with recent precision
measurements of the optical transparency in the visible frequency
range \cite{nair}.

In this Letter we address a different aspect of the phenomenon of the minimal conductivity which has not been discussed before. We begin by pointing out that the minimal conductivity is
essentially a {\it linear response} concept, which needs to be reevaluated in relation to Dirac fermions. For comparison, in a conventional metal the linear
response formalism is reliable as long as the energy
acquired in the external electric field over one period of its
oscillation, $eEv/\omega$, (or the mean free time, if disorder
scattering can not be ignored) must be negligible compared with
the Fermi energy $E_F$ that is typically very large.

In contrast, in clean intrinsic graphene Fermi energy vanishes, $E_F=0$, and the above reasoning fails. To
gain a better understanding of the essential physics, let us take a closer  look
at the Golden rule derivation of the ac conductivity
$\sigma_0$. Fig.~1 illustrates a vertical interband transition between the
two states in the filled lower cone and the empty upper cone, which is responsible for the energy dissipation in electric field.
The matrix element of this transition is of the order of the energy gained in external field, $H_{12}\sim eEv/\omega$.
The number of available electrons in the lower cone with energy $\hbar
\omega/2$ is determined by the density of states which is linear in frequency, $D(\omega) \sim
\omega/v^2\hbar$. Golden rule expression for the energy absorbtion rate
in these transitions is simply, $\dot {\cal E} \sim \omega
|H_{12}|^2 D(\omega) \sim (e^2/\hbar)  E^2$. Recalling the usual Joule heat
expression we arrive at the electric conductivity independent of
frequency and of the order of the conductance quantum.
\begin{figure}[h]
\centerline{\includegraphics[width=47mm,angle=0,clip]{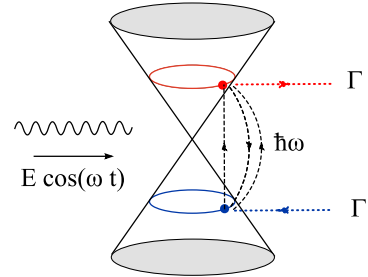}}
\caption{Interband transitions induced by external electric field of
frequency $\omega$. Dashed lines stand for a single transition (short lifetime
$1/\Gamma$)  or multiple Rabi oscillations (long lifetimes
$1/\Gamma$). Dotted lines illustrate energy dissipation from
relaxation of electrons in the upper Dirac cone and holes in the
lower cone.} \label{fig1}
\end{figure}

It is well-known, however, that the first order perturbation theory
fails close to a resonance, i.e.\ when the matrix element becomes of the order of the {\it
detuning} energy, which in our problem is $\hbar \omega-2vp$. Since
available momenta are continuous there are always states with the vanishing
detuning, in which case the inverse {\it lifetime}
$\Gamma$, calculated for a state with the corresponding momentum, must be used as a cut-off. This condition
imposes the restriction upon the strength of electric field, $eE\ll
\hbar \omega \Gamma/v$. As frequency decreases, $\omega \to 0$,
the amplitude of applied field must vanish {\it significantly
faster}, at least as fast as $\sim \omega^2$ \cite{comment_intro}.
Here we concentrate on the response at the same frequency as the frequency of external field, disregarding possible effects of frequency multiplication.
From the above arguments one can expect that the amplitude of induced current will be,
\begin{equation}
\label{current_expansion}
 j =\sigma_0 E\left(1+ C ~\frac{ e^2 E^2
v^2}{\hbar^2 \omega^2 \Gamma^2}+ ... \right),
\end{equation}
and singular at low frequencies. In the present Letter we identify
the leading terms in the series (\ref{current_expansion}) with
the Rabi oscillations between states in the lower and upper cones
with the same momentum.

The above analysis indicates that the lifetime $\Gamma$ plays a central
role in the calculation of ac conductivity. When $\Gamma$ is small, or more precisely, when the dimensionless parameter
\begin{equation}
\label{zeta}
\zeta = \frac{evE}{\hbar \omega\Gamma}
\end{equation}
is large, Rabi oscillations between states in the two cones persist
for a long time and make energy dissipation less effective thus
decreasing the conductivity. In the opposite limit, $\zeta \ll 1$,
fast relaxation ensures that the states in the upper cone are always
empty and that no Rabi oscillations can occur, thus making
first-order perturbation theory applicable. We find that a homogeneous external electric field applied
along the $x$-axis within the plane of graphene
\begin{equation}
{\bf E}(t)={\bf \hat x} E \cos{\omega t},
\end{equation}
induces the electric current which  for arbitrary values of $\zeta$ has the form \cite{comment},
\begin{equation}
\label{result}
{\bf j}= {\bf \hat x}   \frac{2\sigma_0E\cos{\omega t}}{\sqrt{1+\zeta^2}+1}.
\end{equation}
We observe that the minimal conductivity is recovered when $\zeta
\to 0$, however for strong electric fields (or small frequencies),
when $\zeta  \gg 1$, the current {\it saturates} at the value
\begin{equation}
\label{current_satur} j_{max}=\frac{|e|\omega \Gamma}{2 v}.
\end{equation}
This result can be qualitatively visualized as one electron being
transferred per its lifetime $1/\Gamma$ across each strip of graphene that has the
width equal to the electron wavelength $ v/\omega$ corresponding to energy
$\hbar\omega$.

{\it Derivation.} Homogeneous external electric field is most conveniently incorporated in the electron dynamics
with the help of the longitudinal gauge and vector potential
\begin{equation}
{\bf A}(t)=-{\bf \hat x} \frac{c}{\omega}E \sin{\omega t}.
\end{equation}
The Hamiltonian of graphene is the well-known Dirac Hamiltonian in
the pseudospin (sublattice) space
\begin{equation}
\label{ham} \hat H=v{\bm \sigma}\cdot  {\bf p}+\frac{ev}{\omega} \hat
\sigma_x E \sin{\omega t}.
\end{equation}
The crucial observation that allows the solution of the problem
beyond first order perturbation theory is that the momentum ${\bf
p}$ is an integral of motion. The external electric field therefore
mixes only the two states with exactly the same momentum. Denoting by
$a_{\bf p}(t)$ the amplitude to occupy state in the lower cone and
by $b_{\bf p}(t)$ the same for the upper cone we write the total wave
function as,
\begin{eqnarray}
\hat \psi_{\bf p}(t) = \frac{a_{\bf p}(t)}{\sqrt{2}} \left(\begin{array}{r} e^{-i\chi_{\bf p}/2}\\ - e^{i\chi_{\bf p}/2} \end{array} \right)e^{ivpt} \nonumber\\ + \frac{b_{\bf p}(t)}{\sqrt{2}} \left(\begin{array}{r} e^{-i\chi_{\bf p}/2}\\  e^{i\chi_{\bf p}/2} \end{array} \right)e^{-ivpt},
\end{eqnarray}
where $\chi_{\bf p}$ is the angle between the electron momentum
${\bf p}$ and the $x$-axis. Spinor Schr\"odinger equation $i\partial
\hat \psi_{\bf p}/\partial t=\hat H\hat \psi_{\bf p}$ reduces to a pair
of equations (we assume for brevity that $\hbar=1$),
\begin{eqnarray}
\label{amplitudes_eqs}
i\dot a_{\bf p} &=& \frac{evE_0}{\omega} \sin{\omega t}~(i b_{\bf p}\sin\chi_{\bf p} e^{-2ivpt}-a_{\bf p} \cos\chi_{\bf p}),\nonumber\\
i\dot b_{\bf p} &=& \frac{evE_0}{\omega} \sin{\omega t}~(b_{\bf p} \cos\chi_{\bf p}-ia_{\bf p}\sin\chi_{\bf p} e^{2ivpt}).
\end{eqnarray}
In solving Eqs.~(\ref{amplitudes_eqs}) we follow the established
route by keeping only the low-frequency resonant terms, $\sim e^{\pm
i(\omega-2vp)t}$, the approximation known in quantum optics as the
rotating wave approximation.
An additional modification of the equations of motion is needed,
however. In the present form Eqs.~(\ref{amplitudes_eqs}) lack
relaxation and describe undamped Rabi oscillations
leading to the vanishing energy dissipation. To include energy relaxation we take
into account the finite lifetime $1/\Gamma_p$, which due to
electron-hole symmetry is assumed to be the same for electrons in the upper
cone and holes in the lower cone:
\begin{eqnarray}
\label{amplitudes_eqs1}
\dot a_{\bf p} &=& -i\Omega_{\bf p} e^{i(\omega-2vp)t}b_{\bf p}+\frac{\Gamma_p}{2}(1-a_{\bf p}),\nonumber\\
\dot b_{\bf p} &=& -i\Omega_{\bf p} e^{-i(\omega-2vp)t}a_{\bf
p}-\frac{\Gamma_p}{2} b_{\bf p},
\end{eqnarray}
where we introduced the Rabi frequency $\Omega_{\bf p}=
\frac{evE_0}{2\omega} \sin\chi_{\bf p}$. While the relaxation is included into Eqs.~(\ref{amplitudes_eqs1}) in a phenomenological way, it has an advantage of allowing analytical solution. After straightforward calculations,
\begin{eqnarray*}
a_{\bf p}(t)&=& \frac{\Gamma_p^2/4-i\Gamma_p \Delta_p/2}{\Omega^2_{\bf
p}+\Gamma_p^2/4-i\Gamma_p \Delta_p/2} +e^{i\Delta_p t/2-\Gamma_p t/2}
 \nonumber\\&& \times \sum C_\pm e^{\pm it\sqrt{\Omega_p^2+\Delta^2/4}},\\
b_{\bf p}(t)&=& \frac{-i\Gamma_p \Omega_{\bf p}/2}{\Omega^2_{\bf
p}+\Gamma_p^2/4-i\Gamma_p \Delta_p/2}e^{-i\Delta_p t} -e^{i\Delta_p
t/2-\Gamma_p t/2} \nonumber\\ &&\times  \sum C_\pm
\frac{\frac{\Delta_p}{2}\pm \sqrt{\Omega_{\bf
p}^2+\frac{\Delta_p^2}{4}}}{\Omega_{\bf p}}e^{\pm
it\sqrt{\Omega_{\bf p}^2+\Delta_p^2/4}},
\end{eqnarray*}
where $\Delta_p=\omega-2vp$ denotes the detuning between the frequency of electric field and the electron-hole excitation energy. The terms containing
constants $C_+$ and $C_-$ depend on the initial conditions but decay
with time and thus are unimportant for the properties of a
``stationary'' state.

Expectation value of electric current carried by a pair with momentum ${\bf p}$ is given by, $j^x_{{\bf p}}(t) =ev\hat \psi_{\bf p}^\dagger (t) \hat \sigma_x \hat\psi_{\bf p}(t)$. Note that the vector potential term in Eq.~(\ref{ham}) does not modify the expression for current since the former is independent of momentum. We obtain,
\begin{eqnarray}
\label{current_resolved}
j^x_{{\bf p}}(t)= E \frac{e^2 v^2}{8\omega}\frac{\Gamma_p^3 \cos{\omega t}+2\Gamma_p^2\Delta_p \sin{\omega t} }{
\left(\Omega^2_{\bf p}+{\Gamma_p^2}/{4}\right)^2+{\Gamma_p^2} \Delta_p^2/4} \sin^2{\chi_{\bf p}}.
\end{eqnarray}
The total current is found upon summation over all momenta, two
spins directions and both Dirac points, $ {\bf j} =4\sum_{\bf
p}  {\bf j}_{\bf p}$. For the first contribution, $\sim \cos{\omega t}$, which is responsible for energy dissipation, it gives,
\begin{equation}
\label{current_total}
j_x (t)= \cos{\omega t} ~\frac{e^2 E}{2\pi^2\omega}
\int\limits_0^{2\pi} \int\limits_0^\infty \frac{\Gamma_p ~p dp
~\sin^2{\chi} d\chi}{(2p-\frac{\omega}{v})^2+\beta^2(p,\chi)},
\end{equation}
where we introduced the notation
$$
\beta(p,\chi)=\frac{\Gamma_p}{4v}+\frac{e^2 vE^2\sin^2{\chi}}{4
\Gamma_p \omega^2}.
$$
Provided that $\beta (p,\chi) \ll \omega/v$ the integrand is a sharply peaked
function of $p-\omega/2v$ and the integration over $dp$ can be
carried out easily. Two conditions must be satisfied in order to ensure this. The first one is essentially the condition that electrons are well-defined excitations, $\Gamma \ll \omega$. The second condition imposes the restrictions on the upper value of electric field. In terms of the dimensionless parameter defined earlier, Eq.~(\ref{zeta}), this condition implies that $\zeta \ll \omega/\Gamma$.

Integrating the sharply peaked Lorentian over $dp$ and taking the subsequent integration over the angle we obtain,
\begin{equation}
\label{current_final}
j_x (t)= \cos{\omega t} ~\frac{ \omega^2\Gamma^2}{2v^2 E} \left(\sqrt{1+\frac{e^2v^2E^2}{\omega^2 \Gamma^2}}-1 \right),
\end{equation}
which is equivalent to Eq.~(\ref{result}). The lifetime in
Eqs.~(\ref{zeta}-\ref{result}) and (\ref{current_final}) is taken at
the  value of momentum corresponding to frequency $\omega$:
$\Gamma \equiv \Gamma_{\omega/2v}$.

The second term in the  numerator of Eq.~(\ref{current_resolved})
describes the out-of-phase non-dissipative part of the current. This
term is odd in $\Delta_p=\omega-2vp$ and thus vanishes after
integration over the vicinity of resonant momenta. Note that the
contribution of non-resonant momenta to the $\sim \sin{\omega t}$ term though small is not necessarily zero, but is beyond the reach of the
approximations employed in our analysis.

{\it Scattering mechanisms}. The dependence  of the inverse lifetime
$\Gamma(\omega)$ on frequency determines the value of the saturation
current (\ref{current_satur}) and depends on the importance of
various scattering mechanisms that could be present in graphene.

i) In case of short-range impurities (vacancies, non-charged
substitutions) the  electron lifetime in the first Born
approximation is simply proportional to the density of states,
$\Gamma(\omega) \propto D(\omega) \propto \omega$ and the saturation
current grows as the second power of the frequency of electric
field.

ii) When charged impurities dominate, the amplitude of the bare
Coulomb impurity potential is inversely proportional to the electron
momentum, ${2\pi e^2}/q \propto 1/\omega$. Since the screening in
intrinsic graphene does not modify this dependence, the lifetime
$\Gamma (\omega) \propto D(\omega)/\omega^2 \propto 1/\omega$ is
inversely proportional to the frequency \cite{nomura}. Thus, the
value of the saturation current is independent of frequency. In both
cases of neutral and charged impurities $j_{max}$ is inversely
proportional to the concentration of impurities.

iii) Electron-phonon scattering at frequencies above the Debye
frequency is quasi-elastic and follows the same linear dependence
originating from the density of states as is the case for short-range
disorder, $\Gamma \propto 1/\omega$ \cite{park}.

{\it Implications for optical experiments}. Of particular
significance for ongoing experiments \cite{li,kuzmenko,mak,nair} is the
calculation of the response of a graphene layer to an incident
electromagnetic wave in the nonlinear regime. Below we assume graphene
suspended in vacuum (or air) and determine its optical transparency.
\begin{figure}[h]
\centerline{\includegraphics[width=83mm,angle=0,clip]{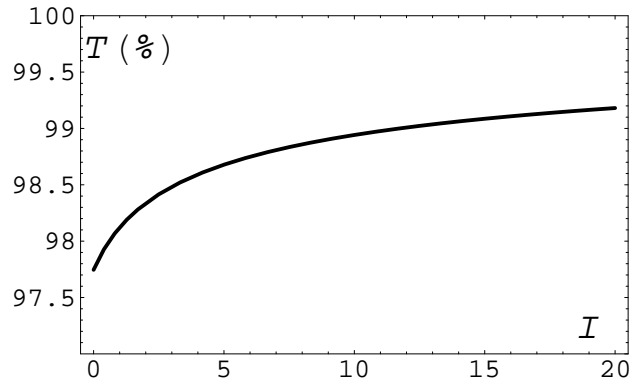}}
\caption{Dependence of the transmission coefficient $T$ of suspended
graphene on the effective intensity of the incident radiation, $I =
(evE_0/\hbar\omega\Gamma)^2 \equiv \zeta_0^2$. At low
intensities it starts from the value of $97.7\%$ corresponding to
the first order perturbation theory but increases sharply with the
intensity $I$, reaching $99\%$ for $\zeta_0 \sim 3$.} \label{fig2}
\end{figure}

The incident radiation is assumed to be propagating along the
negative direction of the $z$-axis and normal to the plane of
graphene ($xy$-plane) with the linear polarization directed along
the $x$-axis. The electric field  consists of the incident  $E_0$
and reflected $E_R$ waves above the graphene flake, and of the
transmitted wave $E_T$ below it
\begin{eqnarray} E=\left\{ \begin{array}{ll} E_0
\cos{(\omega t+\frac{\omega z}{c})}+E_R \cos{(\omega t-\frac{\omega
z}{c})}, & z>0,\\ E_T \cos{(\omega t+\frac{\omega z}{c})}, & z<0.
\end{array}   \right.
\end{eqnarray}
Maxwell's equations with the two-dimensional current provided by graphene sheet at $z=0$ taken
into account has the form,
\begin{equation}
\label{maxwell} \frac{\partial^2 E}{\partial
z^2}-\frac{1}{c^2}\frac{\partial^2 E}{\partial t^2}=-\frac{2\pi
\sigma_0 \omega}{c^2}\frac{E_T\sin{\omega t}}{\sqrt{1+\zeta_T^2}+1}
~\delta (z),
\end{equation}
where $\zeta_T = \frac{evE_T}{\hbar \omega\Gamma}$ corresponds to
the field acting in the graphene flake, $E_T$. The first boundary
condition requires the continuity of the electric field,
$E_0+E_R=E_T$. The second condition is  obtained from integrating
Eq.~(\ref{maxwell}) across the plane of graphene.

Solving for $E_R$ and $E_T$ we obtain the following algebraic
equation for the transmission coefficient $T=E_T^2/E_0^2$,
\begin{equation}
 1+\frac{\pi \alpha}{1+\sqrt{1+T  I}}=\frac{1}{\sqrt{T}},
\end{equation}
where $\alpha=e^2/\hbar c$ stands for the fine structure constant,
and the dimensionless intensity of the incident wave is introduced,
$I = (evE_0/\hbar\omega\Gamma)^2$. Fig.~\ref{fig2} shows the
dependence of the transmission coefficient on the magnitude of the
incident wave. At low intensities, $I\to 0$ the transmission
approaches the linear response value $T=97.7\%$, reported in the
experiment \cite{nair}. As the intensity grows the transmission  initially
increases steeply, but eventually tends much more slowly
 towards $T=1$ according to, $T\approx 1-2\pi \alpha/\sqrt{I}$.

The modern experimental capabilities are well within the necessary
means to resolve the nonlinear transmission, as predicted by Fig.~\ref{fig2}. In
particular, the precision reported in Ref.~\cite{nair} was $0.1\%$.
It is interesting to note the upward trend in $T(\lambda)$ in the
data presented in Ref.~\cite{nair} for different wavelengths
$\lambda$. It is qualitatively reminiscent of the $T(I)$ dependence
predicted in the present Letter, if one assumes short-range
impurities and/or phonons as the primary scattering mechanisms.
Understandably, systematic and more detailed measurements of the
intensity dependence are necessary before any comparisons could be made
with the theory.

{\it Summary.} The conductivity of intrinsic graphene, $e^2/4\hbar$, has been
established to remains constant even when the frequency tends to
zero despite the vanishing of the density of available states,
$D(\omega) \propto \omega \to 0$. This celebrated result originates
in the fact that the matrix element, $evE/\omega$, for the corresponding interband
transition between a filled state in the lower Dirac cone and an
empty state in the upper cone {\it diverges} at low
frequencies. This divergence implies the failure of the first order
perturbation theory and, therefore {\it inapplicability} of the
minimal conductivity at low frequencies for a {\it fixed} strength of
applied electric field. The Golden rule approach is valid only if the magnitude of electric field vanishes at least as $\omega^2$, so that $\zeta \ll 1$. In this paper we extended the theory of electrical conductivity towards finite electric fields for arbitrary values of $\zeta$, as long as $\zeta \ll \omega/\Gamma(\omega)$.
The key observation that allows a solution beyond perturbation
theory is the fact that momenta are conserved since the
important transitions are direct, see Fig.~(\ref{fig1}). The
problem is therefore reduced to strong coupling within simple (and independent of each other)
two-level systems. The electric conduction is largely determined by
those states that are in the vicinity of the resonant transition,
$|\omega-2vp| \sim \Gamma$. When the matrix element for the
transition exceeds the inverse lifetime $1/\Gamma$ multiple
Rabi oscillations develop that greatly reduce the probability of energy
dissipation (Joule heat) and, hence, conductivity. As shown above this must result in the experimentally observable enhancement of the transparency of graphene.

The existing experiments have been performed in the infrared
\cite{li, kuzmenko, mak} and visible light \cite{nair} domains.
Previously, however, nonlinear effects have not been systematically
searched for. We expect that the stronger nonlinear signatures could
be revealed for clean samples, strong intensities of incident
radiation and lower frequencies (possibly infrared).

 Useful discussions with M. Raikh, A. Shytov, and O.
Starykh  are gratefully acknowledged. This work was supported by
DOE, Office of Basic Energy Sciences,
 Award No.~DE-FG02-06ER46313.


\begin{thebibliography}{50}


\bibitem{ludwig}
A.W.W. Ludwig, M.P.A. Fisher, R. Shankar, and G. Grinstein,
Phys. Rev. B {\bf 50}, 7526 (1994).

\bibitem{ryu} In the present paper we consider clean graphene and assume that frequency is larger than the broadening of electron states,
$\omega \gg \Gamma$. In the opposite limit, $\omega \ll \Gamma$, the linear conductivity  has to be determined by means of a self-consistent
theory, see e.g. S. Ryu, C. Mudry, A. Furusaki, and A.W.W. Ludwig, Phys. Rev. B {\bf 75}, 205344
(2007), and is numerically different, $\widetilde \sigma = 2e^2/\pi^2 \hbar$. We leave the nonlinear effects in $\widetilde \sigma$ for future study.

\bibitem{li}Z.Q. Li, E.A. Henriksen, Z. Jiang, Z. Hao, M.C. Martin, P. Kim,
H.L. Stormer, and D.N. Basov, Nature Physics {\bf 4}, 532 (2008).


\bibitem{kuzmenko} A.B. Kuzmenko, E. van Heumen, F. Carbone, and D. van der
Marel,
Phys. Rev. Lett. {\bf 100}, 117401 (2008).

\bibitem{mak} K.F. Mak, M.Y. Sfeir, Y. Wu, C.H. Lui, J.A. Misewich, and T.F.
Heinz, Phys. Rev. Lett. {\bf 101}, 196405 (2008).

\bibitem{2} E.G. Mishchenko, Phys. Rev. Lett. {\bf 98}, 216801 (2007).


\bibitem{sheehy} D.E. Sheehy and J. Schmalian, Phys. Rev. Lett. {\bf 99}, 226803 (2007).

\bibitem{herbut} I.F. Herbut, V. Juricic, and O. Vafek,
Phys. Rev. Lett. {\bf 100}, 046403 (2008).

\bibitem{mishchenko} E.G. Mishchenko, Europhys. Lett. {\bf 83}, 17005 (2008).

\bibitem{sheehy2} D.E. Sheehy and J. Schmalian, ArXiv:0906.5164.


\bibitem{nair} R.R. Nair, P. Blake, A.N. Grigorenko, K.S. Novoselov, T.J.
Booth, T. Stauber, N.M.R. Peres, and A.K. Geim,  Science {\bf 320},
1308 (2008).

\bibitem{comment_intro} For the quasiparticle picture to be meaningful it
is  in any case necessary to have $\Gamma \ll \omega$.

\bibitem{comment} In this paper we concentrate mostly on
the part of the current that is in phase with external electric
field, $j(t)=j \cos{\omega t}$ rather than the out-of-phase
contribution, $\sim \sin{\omega t}$, which does not lead to energy
dissipation. See also explanation below Eq.~(\ref{current_final}).

\bibitem{nomura} K. Nomura and A. H. MacDonald, Phys. Rev. Lett. {\bf 98}, 076602 (2007).

\bibitem{park} C-H. Park, F. Giustino, M.L. Cohen, and S.G. Louie, Phys. Rev. Lett. {\bf 99}, 086804
(2007).
\end{thebibliography}
\end{document}